\documentclass[aps,twocolumn]{revtex4}%
\usepackage{amsmath}
\usepackage{graphicx}
\usepackage{amsfonts}
\usepackage{amssymb}%
\begin{document}
\preprint{HEP/123-qed}
\title{Microwave saturation of the Rydberg states of electrons on helium}
\author{E.Collin, W.Bailey, P.Fozooni, P.G.Frayne, P.Glasson, K.Harrabi, M.J.Lea and G.Papageorgiou}
\affiliation{Royal Holloway, University of London, Egham, Surrey TW20 0EX, England}

\begin{abstract}
We present measurements of the resonant microwave excitation of the Rydberg
energy levels of surface state electrons on superfluid helium. The temperature
dependent linewidth $\gamma(T)$ agrees well with theoretical predictions and
is very small below 300 mK. Absorption saturation and power broadening were
observed as the fraction of electrons in the first excited state was increased
to 0.49, close to the thermal excitation limit of 0.5. The Rabi frequency
$\Omega$ was determined as a function of microwave power. The high values of
the ratio $\Omega/\gamma$ confirm this system as an excellent candidate for
creating qubits.

\end{abstract}


\volumeyear{year}
\volumenumber{number}
\issuenumber{number}
\eid{identifier}
\date{31 July 2002}
\startpage{001}
\endpage{002}
\maketitle

\begin{figure}[th]
%h=here, t=top, b=bottom, p=separate figure page
\par
\begin{center}
\includegraphics[width=1.1\linewidth]{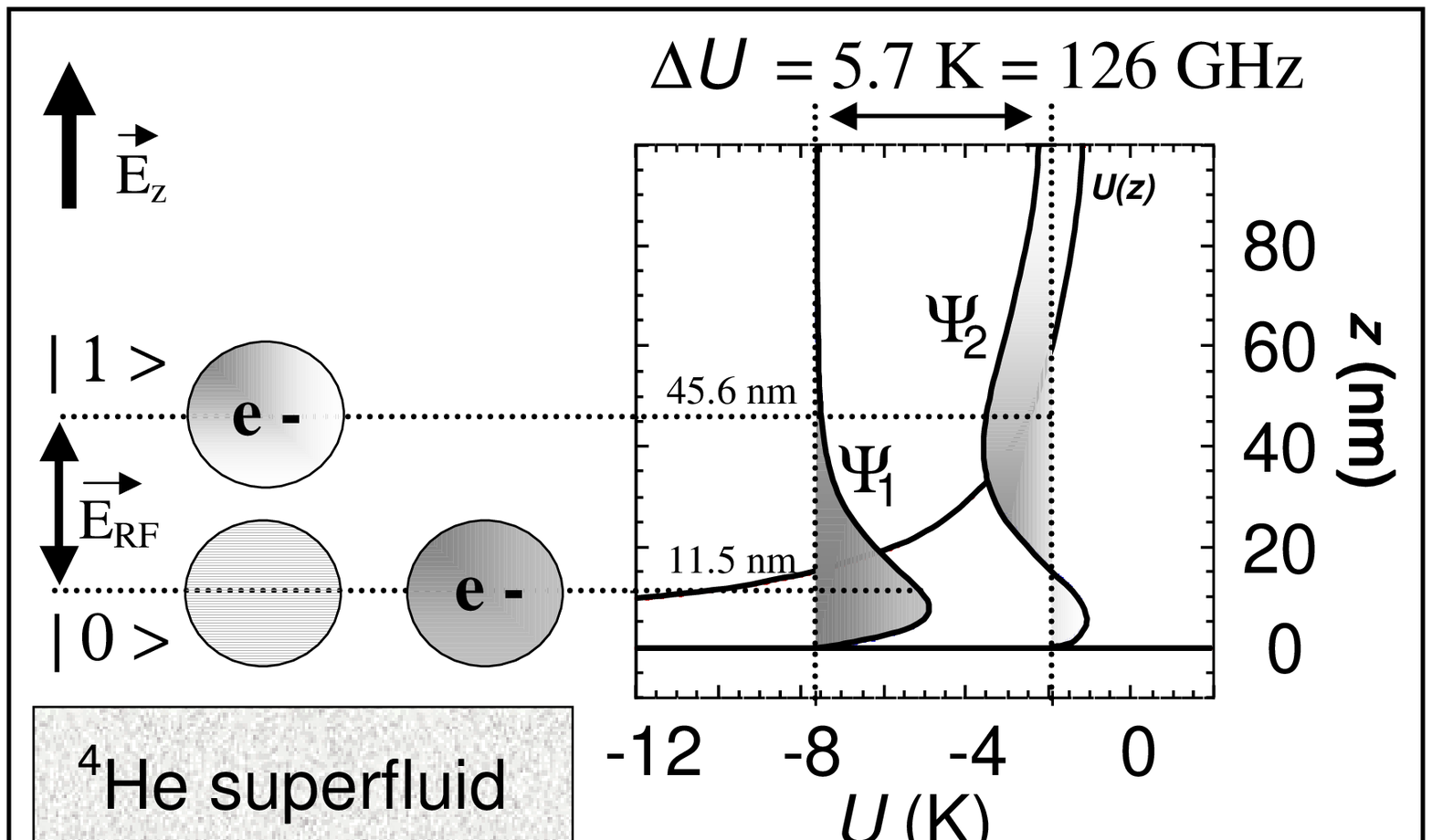}
\end{center}
\caption{The ground ($\left\vert 0\right\rangle $, $m=1$) and first
($\left\vert 1\right\rangle $, $m=2$) excited Rydberg states for electrons on
helium for $E_{z}=0$, showing the Coulomb potential $U(z)$ (solid line) and
the electronic wave functions $\Psi_{1}(z)$ and $\Psi_{2}(z)$. The Stark
effect comes from the different mean heights of the states above the helium
(45.6 nm - 11.5 nm), creating a dipolar moment when $E_{z} \neq0$ is applied.
Transitions from $\left\vert 0\right\rangle $ to $\left\vert 1\right\rangle $
are induced via a vertical microwave electric field $E_{RF}$.}%
\label{figure1}%
\end{figure}

Surface state electrons on liquid helium \cite{Andrei} are attracted by a weak
positive image charge in the helium. This Coulomb potential well for vertical
motion produces a hydrogen-like spectrum, with energy levels $E_{m}%
=-R_{\mathrm{e}}/m^{2}$ where $R_{\mathrm{e}}$ is the effective Rydberg energy
(0.67 meV) and $m(\geq1)$ is the quantum number, shown schematically in Fig.
\ref{figure1}. These Rydberg states were first observed by Grimes \textit{et
al.} \cite{Grimes} who measured the resonant frequency $f_{1m}$ and the
linewidth $\gamma(T)$ above 1.2 K for transitions from the ground state to the
excited states $m=2,3$. A key feature is a linear Stark effect, due to
asymmetry of the wavefunctions, so that the resonant frequency can be tuned
with a vertical electric field $E_{z}$. Their measurements of $f_{1m}$ versus
$E_{z}$ showed excellent agreement with theoretical calculations from
$f_{12}=125.9$ GHz in zero holding field up to 220 GHz for $E_{z}=17.5$ kV/m.
Lambert and Richards \cite{Lambert1} extended the frequency range up to 765
GHz using a far-infrared laser. Edel'man and later Volodin and Edel'man
\cite{Edel'man} indirectly probed these Rydberg states by measuring the
photo-conductivity, or the change in the electron mobility, when the electrons
were excited by incident microwaves. They estimated that the
temperature-dependent linewidth $\gamma(T)<30$ MHz at 0.4 K, consistent with
the theory of Ando \cite{Ando}.

Interest in these states has now been rekindled by the suggestion of Platzman
and Dykman \cite{Phil and Mark} that electrons on helium could be used as
electronic qubits, with the ground and the first excited states representing
$\left\vert 0\right\rangle $ and $\left\vert 1\right\rangle $ respectively.
The qubits would be controlled using resonant microwave excitation. The
potential is anharmonic so that the two lowest states are an excellent
approximation to a two-level system with minimal coupling to higher levels.
The formalism is similar to nuclear magnetic resonance (NMR), governed by the
optical Bloch equations \cite{Loudon}. The linewidth is related to the NMR
relaxation rate $1/T_{1}$ and the decoherence rate $1/T_{2}$, describing the
quantum system \cite{Abragam}. In principle, a surface state electron quantum
computer would be similar to the NMR quantum computers recently developed
\cite{NMRQC1} \cite{NMRQC2}. The potential advantages come from the
exceptional properties of the electronic system itself and the scalability of
custom-designed qubits. The results presented here were obtained as part of an
experimental programme to develop such a system \cite{LeaQC}.

In particular, we report the first direct measurements below 1.2 K of the
microwave absorption between the Rydberg states of surface state electrons on
helium. Two key aspects for qubits are the linewidth of the resonant microwave
absorption, which must be small to limit decoherence, and the ability to
excite a high fraction of electrons into the excited state. We have measured
the temperature dependent linewidth $\gamma(T)$ at 189.6 GHz, with values of
$\gamma(T)$ over two orders of magnitude smaller than previous direct
measurements, into the ripplon scattering regime. We have made the first
measurements of saturation and power broadening of the absorption line in this
system, due to the finite occupancy of the first excited state. High levels of
electron excitation were achieved, very close to the limit for thermal
equilibrium. The Rabi frequency was determined as a function of microwave power.

Power from a Gunn diode oscillator \cite{Radiometer} was passed through a
doubler (5 mW maximum output from 165 to 195 GHz) and transmitted down
overmoded waveguide, through thermal filters, into an experimental cell on a
dilution refrigerator. The frequency of the Gunn oscillator $f=\omega/2\pi$
was phase-locked to a 10 MHz quartz crystal resonator. Higher frequencies, up
to 260 GHz, were obtained from a carcinotron source. The electrons were held
above the liquid helium between capacitor plates 2 mm apart which formed a
flat cylindrical cavity, 52 mm diameter. The microwaves were polarised
vertically by a wire grid on the cavity input port and propagated horizontally
through the cell to a low temperature InSb Putley bolometer \cite{QMC}.
The\ vertical holding field $E_{z}$ was swept by varying the potential
difference $V_{z}$ between the capacitor plates. Two techniques were used to
measure the absorption linewidth. First, the holding field was sine-wave
modulated (typically by 10 mV rms) at 5 kHz and the differential absorption
signal measured by the Putley detector using a lock-in amplifier. This was
integrated numerically to obtain the absorption lineshape. In the second
method a larger amplitude (3 V) square wave pulse was applied, such that each
half-cycle alternately sampled the resonant absorption or a region of very
small absorption. The two methods gave identical lineshapes.

\begin{figure}[th]
%h=here, t=top, b=bottom, p=separate figure page
\par
\begin{center}
\includegraphics[width=1.2\linewidth]{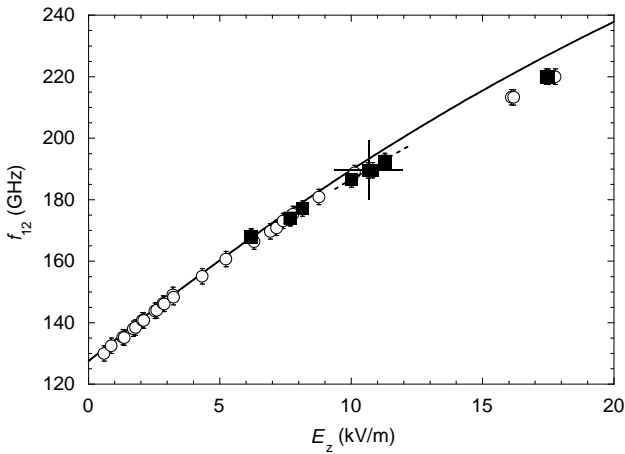}
\end{center}
\caption{Resonant frequency $f_{12}$\ versus $E_{z}$ at 1 K, for low microwave
powers (full squares). The empty circles are from \cite{Grimes}. The solid
line shows the theoretical calculations of \cite{Grimes}. The cross represents
the frequency (189.6 GHz) we have chosen for our temperature/powers
characterizations. The dashed line is the slope of the experimentally measured
curve at this point.}%
\label{figure2}%
\end{figure}

The resonant frequency $f_{12}=\omega_{12}/2\pi$ versus $E_{z}$\ is shown in
Fig. \ref{figure2}, in excellent agreement with previous experiments
\cite{Grimes}\cite{Lambert1}. The integrated lineshape above 1 K is close to
the Lorentzian lineshape%

\begin{equation}
L(\delta)=\frac{\gamma/\pi}{\delta^{2}+\gamma^{2}} \label{Eq(1)}%
\end{equation}
where $\delta=\omega-\omega_{12}$. The half-width $\gamma$ was measured in
terms of the voltage $V_{z}$. The conversion factor to frequency is 2.44
GHz/Volt at 189.6 GHz (the slope of Fig. \ref{figure2}). Below 1 K, the
linewidth decreases rapidly and becomes dominated by the inhomogeneous
linewidth (typically 50 MHz) due to variations in the vertical holding field,
and hence the resonant frequency, across the electron sheet. The inhomogeneous
absorption $\alpha_{0}(V_{z})$ of the cell was measured at low temperatures,
where the intrinsic linewidth is small. As the temperature increased the
experimental absorption lineshape $\alpha(V_{z},T)$ changed and broadened. The
temperature dependent part was obtained by convoluting a Lorentzian
absorption, half-width $\gamma=\gamma(T)$, with the inhomogeneous low
temperature absorption, using $\gamma$ as a best-fit parameter. The fit of the
convoluted lineshape to experiment was excellent at all temperatures,
confirming that the intrinsic temperature dependent lineshape is Lorentzian.
The temperature dependent contribution was less than the error bars below 300
mK, which was our reference temperature.

\begin{figure}[th]
%h=here, t=top, b=bottom, p=separate figure page
\par
\begin{center}
\includegraphics[width=1.2\linewidth]{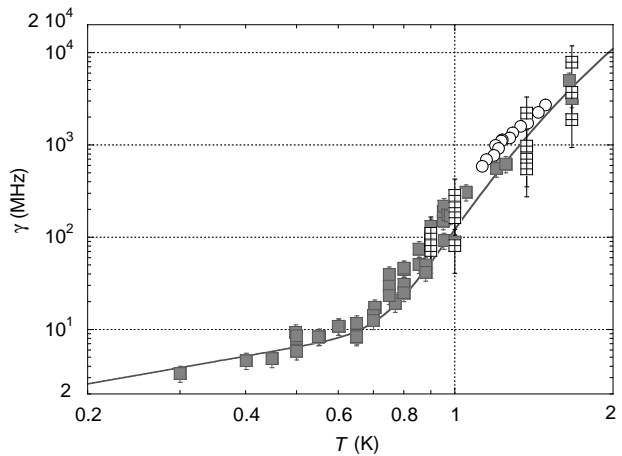}
\end{center}
\caption{Linewidth $\gamma(T)$ (measured as the half-height-half-width) versus
temperature for 189.6 GHz. Full squares are the low power measurements,
obtained as explained in the text. The circles are the data from
\cite{Grimes}. The crossed squares are our results extracted from the power
dependency (see text). The full line is Ando's theory \cite{Ando}.}%
\label{figure3}%
\end{figure}

By analyzing the lineshape in this way, we measured the temperature dependent
half-width $\gamma(T)$ as shown in Fig. \ref{figure3}.\ \ Above 1 K,
scattering from $^{\text{4}}$He vapor atoms dominates and is proportional to
the vapor pressure, while below 1 K, the scattering is from surface waves
(ripplons). The theory by Ando \cite{Ando} gives%

\begin{equation}
\gamma(T)=AT+BN_{\mathrm{gas}} \label{Ando}%
\end{equation}
where the first term is due to ripplon scattering and $N_{\mathrm{gas}}$
$\propto T^{3/2}\exp(-7.17/T)$\ is the number density of $^{\text{4}}$He vapor
atoms. The coefficients $A$ and $B$ depend on the holding field $E_{z}$. The
reference value of $\gamma(T)$ at 300 mK was determined by a linear fit of the
data below 0.7 K, following Eq. (\ref{Ando}).

Both inelastic and elastic collisions contribute to the linewidth. Inelastic
collisions produce decay, or relaxation, of the excited state with a lifetime
$\tau=1/2\gamma_{\mathrm{in}}$ \cite{Gamma} (the radiative lifetime is
estimated to be very long, $\sim0.1$ s, in this system) while elastic
collisions produce fluctuations in the energy levels and hence a linewidth
$\gamma_{\mathrm{el}}$ (and also decoherence). The total Lorentzian half-width
$\gamma=$ $\gamma_{\mathrm{in}}+\gamma_{\mathrm{el}}$. The new experimental
measurements cover three decades in $\gamma(T)$. They are in good agreement
with previous data above 1.2 K, and lie close to the theoretical values shown
by the solid line in Fig. \ref{figure3} \cite{Ando}.

\begin{figure}[th]
%h=here, t=top, b=bottom, p=separate figure page
\par
\begin{center}
\includegraphics[width=1.2\linewidth]{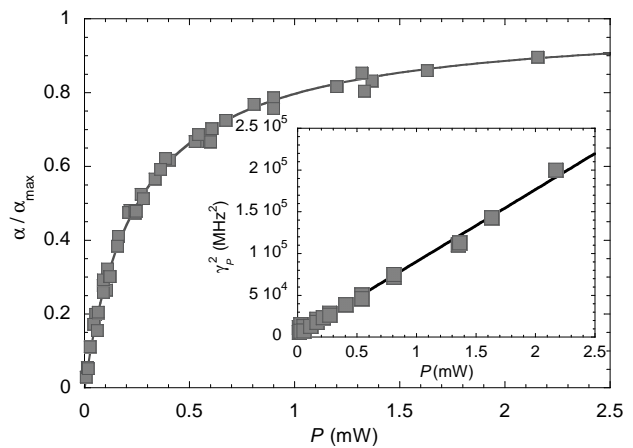}
\end{center}
\caption{Absorption saturation data $\alpha(P)/\alpha_{\max}$ at 0.9 K for
189.6 GHz, normalized to the maximum absorption. The input power $P$ is
measured at room temperature. The line is a fit explained in the text. Inset:
Power broadening data $\gamma_{P}^{2}$ versus $P$, where the inhomogeneous
broadening has been subtracted.}%
\label{figure4}%
\end{figure}

Experimentally the linewidth is independent of the microwave power, at low
powers. As the power increases, the absorption line broadens and saturates.
The saturation absorption $\alpha(P)$ at resonance is shown in Fig.
\ref{figure4}, plotted versus the microwave input power, as measured by a
power meter at the top of the cryostat. The transmission loss from the source
to the detector was typically 20 - 30 dB. At low powers $\alpha=aP$ but it
then saturates as $\alpha(P)=aP/(1+aP/\alpha_{\max})$. Saturation is due to
the finite occupancy of the excited state. The optical Bloch equations reduce
to the rate equation \cite{Loudon} for the fractional occupancies of the
ground and excited states, $\rho_{1}$ and $\rho_{2}=1-\rho_{1}$%

\begin{equation}
\frac{\mathrm{d}\rho_{2}}{\mathrm{d}t}=r(\rho_{1}-\rho_{2})-\frac{\rho_{2}%
}{\tau}=r(1-2\rho_{2})-\frac{\rho_{2}}{\tau}%
\end{equation}
where $r$ is the rate for stimulated absorption and emission which is balanced
by spontaneous emission. In dynamic equilibrium, $\rho_{2}=r\tau/(1+2r\tau)$
with a saturation value of 0.5. This effect directly leads to power saturation
and power broadening. The excitation rate can be expressed in terms of the
Rabi frequency $\Omega=eE_{\mathrm{RF}}z_{12}/\hbar$ where $E_{\mathrm{RF}}$
is the microwave field amplitude and $z_{12}$ is the electric dipole length
for the transition. The excitation rate $r=0.5\Omega^{2}\gamma/(\delta
^{2}+\gamma^{2})$ \cite{Loudon}. Hence the power absorption $\alpha=N\rho
_{2}/\tau$ is%

\begin{equation}
\alpha=\frac{\frac{1}{2}N\gamma\Omega^{2}}{\delta^{2}+\gamma^{2}+\gamma
\tau\Omega^{2}} \label{alpha}%
\end{equation}
where $N$ is the number of electrons. This has the form shown experimentally
in Fig. \ref{figure4} with $\alpha_{\max}=N/2\tau=N\gamma_{\mathrm{in}}$. The
Lorentzian linewidth is now power dependent with $\gamma_{P}^{2}=\gamma
^{2}+\gamma\tau\Omega^{2}=\gamma^{2}+bP$ where $P\propto E_{\mathrm{RF}}%
^{2}\propto\Omega^{2}$ is the incident microwave power. Power broadening is a
direct measure of the Rabi frequency.

Fig. \ref{figure4} shows the experimental power dependent linewidth (with the
inhomogeneous linewidth subtracted), plotted as $\gamma_{P}^{2}$ versus $P$,
confirming the theoretical expression for power broadening. The intercept
gives the low power value for $\gamma(T)$ as plotted in Fig. \ref{figure3}.

The fractional occupancy $\rho_{2}=0.5(1-(\gamma/\gamma_{P})^{2})$ of the
excited state can be obtained directly from the power broadening and has a
maximum value of 0.489 for the data in Fig. \ref{figure4}, very close to the
thermal equilibrium limit of 0.5. The Rabi frequency and the microwave
electric field amplitude can also be obtained from the power broadening term
$\Omega\sqrt{\gamma\tau}$ \cite{Gamma}\ which has a maximum value of 440 MHz
for the data in Fig. \ref{figure4}. The dimensionless factor $\gamma\tau$
depends on the scattering and decay mechanisms producing the finite linewidth.
Variational wavefunctions for the ground state $\Psi_{1}(z)$\ and first
excited state $\Psi_{2}(z)$\ were calculated as a function of the electric
holding field $E_{z}$ \cite{Ando}. Above 0.7 K, the linewidth is primarily due
to electron scattering from $^{\mathrm{4}}$He vapor atoms which behave as
point, or $\delta$-function, scattering centres. In this case, we find
theoretically that $\gamma\tau=7.1$ for $E_{z}=0$, decreasing to 2.9 for
$E_{z}=1.0\times10^{4}$ V/m, corresponding to a resonant frequency of 190 GHz.
This would give a maximum value for $\Omega=260$ MHz for the data in Fig.
\ref{figure4}. The dipole length for excitation $z_{12}=\left\vert \langle
\Psi_{1}^{\ast}(z)\left\vert z\right\vert \Psi_{2}(z)\rangle\right\vert $
increases from 4.2 nm in zero pressing field to 5.1 nm at a resonant frequency
of 190 GHz. The corresponding maximum microwave electric field amplitude would
be 208 V/m. The power dependent data was also analysed, using Eq.
(\ref{alpha}), to obtain values of $\gamma(T)$ above 0.9 K, as shown in Fig.
\ref{figure3}, in good agreement with the low power data.

The total saturation power absorbed by the electrons is estimated as 70 nW at
0.9 K. No heating effects were apparent at this temperature, though these may
become important at lower temperatures.

This analysis assumes that the response is independent of the electron density
$n$. The strong electron-electron Coulomb interactions will not contribute to
the linewidth directly (Kohn theorem) though they can mediate the scattering
by other mechanisms, as in electronic transport in a magnetic field
\cite{Magnetoconductivity}. However, the electron-electron interactions will
change the energies of the Rydberg states and hence the resonant frequency.
Indeed, a basic idea for surface-state qubits is that the excitation of one
electron will change the resonant frequency of its neighbor, via the Coulomb
coupling \cite{Phil and Mark}. For a typical electron density of
$n=0.18\times10^{12}$ m$^{-2}$ and an inter-electron distance of 2.5
$\mu\mathrm{m}$, the Coulomb shift is about $12$ MHz for nearest neighbors.
This introduces a non-linear term into Eq. (\ref{alpha}) which becomes
significant as $\gamma(T)$ becomes less than the Coulomb shift below 0.8 K
\cite{Coulomb}. Controlling these effects for interacting qubits is an overall
objective. The power dependence in this region also changes, due to a change
in the factor $\gamma\tau$ as the scattering mechanism changes.

Significant conclusions can be drawn concerning the use of these states as
qubits for quantum processing. First, we have experimentally achieved the
microwave field amplitudes and Rabi frequencies which were postulated by
Platzman and Dykman \cite{Phil and Mark} in their qubit proposal. The Rabi
frequency would represent the clock frequency for qubits. Secondly, we have
shown that the temperature dependent linewidth $\gamma(T)$ is indeed small at
low temperatures, at least on bulk helium. Kirichek et al. \cite{Kirichek}
recently measured the scattering rate for surface state electrons on both
$^{3}$He and $^{4}$He from the linewidth of electron plasma resonances, down
to 10 mK. On $^{4}$He, they found excellent agreement with the theory of
electron-ripplon scattering. For transport measurements, the scattering rate
remains finite at the lowest temperatures. However, the microwave linewidth
$\gamma(T)$ decreases linearly with temperature \cite{Ando}. In our
experiments at 100 mK, $f_{12}/\gamma\simeq2\times10^{5}$ while $\Omega
/\gamma\approx300$. This demonstrates the high quality of the Rydberg states
in this system. The small single ripplon-induced-decay would be further
suppressed by localising electrons in traps (as required for qubits) and by
the application of a perpendicular magnetic field \cite{Phil and Mark}. The
present experiment gives grounds for optimism for the use of these states in
electronic qubits, though much remains to be done.

We thank M.I.Dykman, P.M.Platzman and J.Singleton for discussions and
F.Greenough, A.K.Betts and others for technical support. The work was
supported by the EPSRC, the EU, INTAS and Royal Holloway, University of London.


\begin{references}
\bibitem{Gamma} Experimentally, we plot $\gamma$ in frequency units (MHz). In
the theoretical expression for $\gamma\tau$,  $\gamma$ is an angular frequency
(rad/s).
\end{references}
\end{document}